\documentclass[twocolumn,prl,aps,showpacs,preor==]{revtex4-1}
\pdfoutput=1 \usepackage{preprintcover}
\RequirePackage[colorlinks,breaklinks,pdftitle={ATLAS
draft},pdfauthor={The ATLAS Collaboration}]{hyperref}
\hypersetup{linkcolor=blue,citecolor=blue,filecolor=black,urlcolor=blue}
\usepackage{graphicx} \usepackage{dcolumn} \usepackage{multirow}
\usepackage{amssymb} \usepackage{amsmath} \usepackage{slashed}
\usepackage{atlasphysics} \usepackage{hyperref} \usepackage{pifont}
\usepackage[usenames,dvipsnames]{xcolor}

\newcommand{\ANpart}{\mbox{$\langle N_{\mathrm{part}}\rangle$}}

\newcommand{\Nevt}{\mbox{$N_{\mathrm{evt}}$}}

\newcommand{\pp}{\mbox{$pp$}} 
\newcommand{\PbPb}{\mbox{Pb+Pb}}

\newcommand{\TAA}{\mbox{$T_{\mathrm{AA}}$}}
\newcommand{\ATAA}{\mbox{$\langle T_{\mathrm{AA}} \rangle $}}
\newcommand{\RAA}{\mbox{$R_{\mathrm{AA}}$}}

\newcommand{\sqrts}{\mbox{$\sqrt{s}$}}
\newcommand{\sqrtsnn}{\mbox{$\sqrt{s_{\mathrm{NN}}}$}}

\newcommand{\kt}{\mbox{$k_t$}}

  \newcommand{\RFour}{\mbox{$R = 0.4$}}

\newcommand{\ETfcal}{\mbox{$\Sigma E_{\mathrm{T}}^{\mathrm{FCal}}$}}

\newcommand{\invnb}{\mbox{$\mathrm{nb}^{-1}$}}
\newcommand{\invpb}{\mbox{$\mathrm{pb}^{-1}$}}

\newcommand{\Dphi}{\mbox{$\Delta \phi$}}
\newcommand{\absy}{\mbox{$|y|$}}

\newcommand{\percent}{\%}
\newcommand{\LumiUncert}{3.1\percent}
\newcommand{\papertype}{Letter}
\newcommand{\papertitle}{Measurements of the Nuclear Modification
Factor for Jets in \PbPb\ Collisions at $\sqrtsnn=2.76$~\TeV\ with the
ATLAS Detector} 
\newcommand{\myabstract}{Measurements of inclusive jet
production are performed in \pp\ and \PbPb\ collisions at
$\sqrtsnn=2.76$~\TeV\ with the ATLAS detector at the LHC,
corresponding to integrated luminosities of 4.0~\invpb\ and
0.14~\invnb, respectively. The jets are identified with the anti-\kt\
algorithm with $R=0.4$, and the spectra are measured over the
kinematic range of jet transverse momentum $32 < \pt < 500$~\GeV, and
absolute rapidity $|y| < 2.1$ and as a function of collision
centrality. The nuclear modification factor, \RAA, is evaluated and
jets are found to be suppressed by approximately a factor of two in
central collisions compared to \pp\ collisions. The \RAA\ shows a
slight increase with \pt\ and no significant variation with rapidity.}

\PreprintCoverPaperTitle{\papertitle}
\PreprintIdNumber{CERN-PH-EP-2014-172}
\PreprintCoverAbstract{\myabstract}
\PreprintJournalName{Physical Review Letters}
\begin{document}
 \preprint{APS/123-QED}
\title{\papertitle}
\author{ATLAS Collaboration}
\begin{abstract}
\myabstract
\end{abstract}
\maketitle 

Relativistic heavy-ion collisions at the LHC produce a medium of
strongly interacting nuclear matter composed of deconfined color
charges
\cite{Adcox:2004mh,Adams:2005dq,Back:2004je,Arsene:2004fa}. Hard
scattering processes occurring in these collisions produce high
transverse momentum (\pt) partons that propagate through the medium
and lose energy, resulting in the phenomenon of ``jet quenching.'' The
partonic energy loss can be probed through measurements of the
suppression of jet production rates. The effects of energy loss have
been observed through the suppression of single
hadrons~\cite{Adler:2003qi,Adams:2003kv,Arsene:2003yk,Back:2004ra,CMS:2012aa,Aamodt:2010jd,Abelev:2012hxa}
and jets constructed from charged particles~\cite{Abelev:2013kqa}.
ATLAS has previously reported measurements with fully reconstructed
jets~\cite{Aad:2012vca} by comparing the jet yields in central
collisions, where the colliding nuclei have large overlap, to the
yields in peripheral collisions. Those results indicate that the rate
of jets in \PbPb\ collisions is suppressed by a factor of
approximately two in central collisions relative to peripheral
collisions. A more sensitive probe of energy loss is provided by
measurements of the suppression relative to \pp\ collisions, where
there are no quenching effects.

The magnitude of the suppression is expected to depend on both the
\pt\ dependence of the energy loss as well as the shape of the initial
jet production \pt\ spectrum~\cite{Adcox:2004mh}. This spectrum
becomes increasingly steep at larger values of the jet rapidity
\cite{Aad:2011fc}. Thus measurements of jet suppression for jets in
different intervals of rapidity provide complementary information
about the energy loss. Additionally, parton showers initiated by
quarks may be quenched differently than gluons~\cite{Baier:2000mf},
and the fraction of quark-initiated jets is expected to increase with
rapidity.

Hard scattering rates are enhanced in more central collisions; the
larger overlap results in a higher integrated luminosity of partons
able to participate in hard scattering processes, and these hard
scattering rates are expected to be proportional to the nuclear
overlap function, \TAA. The suppression is quantified by the nuclear
modification factor
\begin{equation*}
\RAA=\frac{ \left.\frac{1}{\Nevt}
\frac{d^2 N_{\mathrm{jet}}}{d\pt d y} \right
|_{\mathrm{central}}} {\ATAA \, \frac{d^2
  \sigma_{\mathrm{jet}}^{pp}}{d\pt d y} }\,.
\label{eqn:raa_def}
\end{equation*}

This \papertype\ presents measurements of the inclusive jet \RAA\ in
\PbPb\ collisions at a nucleon--nucleon center-of-mass energy of
$\sqrtsnn = 2.76$~\TeV. It utilizes \PbPb\ data collected during 2011
corresponding to an integrated luminosity of 0.14~\invnb\ as well as
data from \pp\ collisions recorded during 2013 at the same
center-of-mass energy corresponding to 4.0 \invpb. Results are
presented for jets reconstructed in the calorimeter with the anti-\kt\
jet-finding algorithm~\cite{Cacciari:2008gp} with jet radius parameter
$R=0.4$. The contribution of the underlying event (UE) to each jet,
assumed to be uncorrelated and additive, was subtracted on a per-jet
basis.

The measurements presented here were performed with the ATLAS
calorimeter, inner detector, trigger, and data acquisition
systems~\cite{atlasfootnote,Aad:2008zzm}. The calorimeter system
consists of a liquid argon (LAr) electromagnetic (EM) calorimeter
($|\eta|<3.2$), a steel-scintillator sampling hadronic calorimeter
($|\eta| < 1.7$), a LAr hadronic calorimeter ($1.5 < |\eta| < 3.2$),
and a forward calorimeter (FCal) ($3.2 < |\eta| <
4.9$). Charged-particle tracks were measured over the range
$|\eta|<2.5$ using the inner detector~\cite{Aad:2010bx}, which is
composed of silicon pixel detectors in the innermost layers, followed
by silicon microstrip detectors and a straw-tube transition-radiation
tracker ($|\eta| < 2.0$), all immersed in a 2~T axial magnetic
field. The zero-degree calorimeters (ZDCs) are located symmetrically
at $z = \pm 140$~m and cover $|\eta| > 8.3$. A ZDC coincidence trigger
was defined by requiring a signal consistent with one or more neutrons
in each of the calorimeters.

The \pp\ events used in the analysis were selected using the ATLAS jet
trigger~\cite{Aad:2012xs} with multiple values of the trigger \pt\
thresholds. During \pp\ data taking, the average number of \pp\
interactions per bunch crossing (pile-up) varied from 0.3 to 0.6.  The
\pp\ events were required to contain at least one primary vertex,
reconstructed from at least two tracks, and jets originating from all
such vertices were included in the cross section measurement.

Data from \PbPb\ collisions were recorded using either a minimum-bias
trigger or a jet trigger. The minimum-bias trigger, formed from the
logical OR of triggers based on a ZDC coincidence or total transverse
energy in the event, is fully efficient in the range of centralities
presented here. The jet trigger identified jets by applying the
anti-\kt\ algorithm with $R=0.2$ with a UE subtraction procedure
similar to that applied in the offline analysis. The jet trigger
selected events having at least one jet with transverse energy $\ET >
20$~\GeV\ at the electromagnetic scale ~\cite{Aad:2011he}. Event
selection and background rejection criteria were
applied~\cite{ATLAS:2011ah} yielding 53 million and 14 million events
in the minimum-bias and jet-triggered samples, respectively.

The centrality of \PbPb\ collisions was characterized by \ETfcal, the
total transverse energy measured in the FCal~\cite{ATLAS:2011ah}. The
centrality intervals were defined according to successive percentiles
of the \ETfcal\ distribution ordered from the most central (highest
\ETfcal) to the most peripheral collisions. A Glauber model analysis
of the \ETfcal\ distribution was used to evaluate the \ATAA\ and the
number of nucleons participating in the collision, \ANpart, in each
centrality
interval~\cite{Miller:2007ri,Alver:2008aq,ATLAS:2011ah}. The
centrality intervals used in this measurement are indicated in
Table~\ref{tab:centrality_table} along with the values of \ATAA\ and
\ANpart\ for those intervals.

\begin{table} 
\centering
\begin{tabular}{|r@{--}l| | r@{$\,\pm\,$}l|r@{$\,\pm\,$}l|}
\hline \multicolumn{2}{|c||}{Centrality [\percent]} &
\multicolumn{2}{c|}{\ATAA\ [$\mathrm{mb}^{-1}$]} &
\multicolumn{2}{c|}{\ANpart} \\ \hline
 0 & 10   & 23.45 & 0.37  &356.2 & 2.5 \\ \hline             
 10 & 20 &14.43 & 0.30      &260.7 & 3.6 \\ \hline             
 20 & 30 &8.73 & 0.26      &186.4 & 3.9 \\ \hline             
 30 & 40 &5.04 & 0.22      &129.3 & 3.8 \\ \hline             
 40 & 50 &2.7 & 0.17       &85.6 & 	3.6 \\ \hline      
 50 & 60 &1.33 & 0.12      &53.0 & 	3.1 \\ \hline      
 60 & 70 &0.59 & 0.07      &30.1 & 	2.5 \\ \hline      
 70 & 80 &0.24 & 0.04      &15.1 & 	1.7 \\ \hline  \hline
 0 & 1 & 29.04 & 0.46      &400.1 & 1.3 \\ \hline             
 1 & 5 & 25.62 & 0.40     &377.6 & 2.2 \\ \hline             
 5 & 10 & 20.59 & 0.34    &330.3 & 3.0 \\ \hline             
 60 & 80 & 0.41 & 0.05    &22.6 & 	2.1 \\ \hline  
\end{tabular}
\caption{The \ATAA\ and \ANpart\ values and their uncertainties in
  each centrality bin.}
\label{tab:centrality_table}
\end{table}

The jet reconstruction and UE subtraction procedures described in
Ref.~\cite{Aad:2012vca} were applied to both \pp\ and \PbPb\ data. The
anti-\kt\ algorithm was applied to logical towers with segmentation
$\Delta\eta\times\Delta\phi=0.1\times0.1$ formed from energy deposits
in the calorimeter. An iterative procedure was used to obtain an
event-by-event estimate of the average \eta-dependent UE energy
density while excluding actual jets from that estimate. The jet
kinematics were obtained by subtracting the UE energy from the towers
within the jet. Following reconstruction, the jet energies were
corrected for the calorimeter energy response using the procedure
described in Ref.~\cite{Aad:2014bia}.

In addition to the calorimetric jets, ``track jets'' were
reconstructed by applying the anti-\kt\ algorithm with \RFour\ to
charged particles with $\pt > 4$~\GeV. In the \PbPb\ analysis, the
track jets were used in conjunction with electromagnetic clusters to
exclude the contribution to the jet yield from UE fluctuations of soft
particles incorrectly interpreted as calorimetric
jets~\cite{Aad:2012vca}. The jets were required to be within $\Delta R
=\sqrt{(\Delta\eta)^2+(\Dphi)^2} =0.2$ of a track jet with $\pT
>7$~\GeV\ or an electromagnetic cluster with $\pt > 8$~\GeV.

The performance of the jet reconstruction in \PbPb\ collisions was
evaluated using the GEANT4-simulated detector response
~\cite{Agostinelli:2002hh,Aad:2010ah} in a Monte Carlo (MC) sample of
\pp\ hard scattering events at $\sqrts = 2.76$~\TeV. The events were
produced with the \textsc{Pythia} event generator
~\cite{Sjostrand:2006za} version 6.423 with parameters chosen
according to the so-called AUET2B tune~\cite{mctunes} and overlaid
with minimum-bias \PbPb\ collisions recorded by ATLAS during the same
data-taking period as the data used in the analysis. Thus the MC
sample contains a UE contribution that is identical in all respects to
the data. A separate \textsc{Pythia} sample was produced for the
analysis of the \pp\ data with the detector simulation adjusted to
match the conditions during the \pp\ data taking including
pile-up. Additional MC samples were used in evaluations of the jet
energy scale (JES) uncertainty. The \textsc{Pyquen}
generator~\cite{Lokhtin:2005px}, which applies medium-induced energy
loss to parton showers produced by \textsc{Pythia}, was used to
generate a sample of jets with fragmentation functions that differ
from those in the nominal \textsc{Pythia} sample in a fashion
consistent with measurements of fragmentation functions in quenched
jets~\cite{Chatrchyan:2012gw,Chatrchyan:2014ava,Aad:2014wha}.
\begin{figure}[t]
  \centering
  \includegraphics[width=0.45\textwidth]{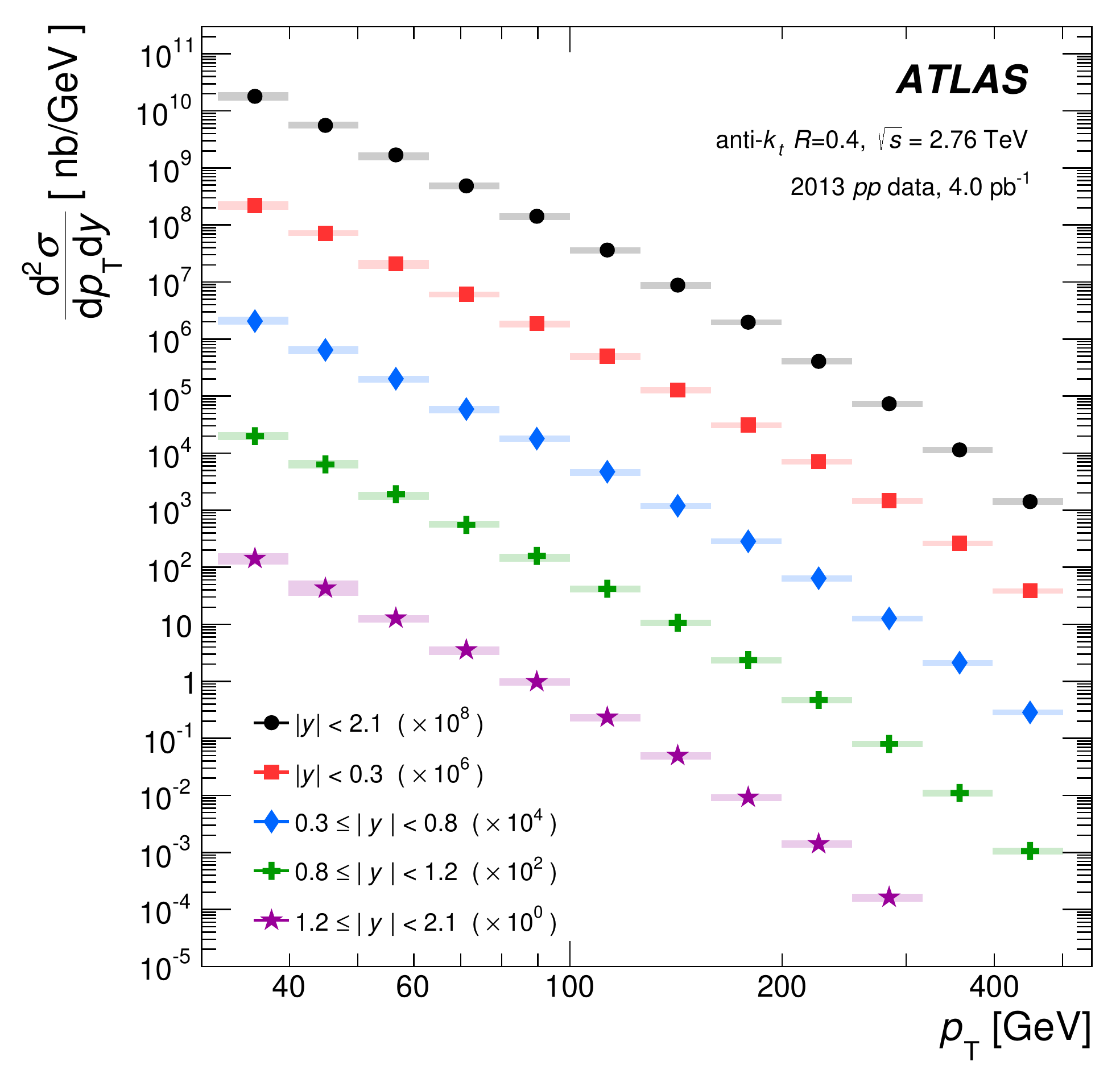}
  \caption{
The double differential jet cross section in \pp\ collisions as a function of \pt\ in
different rapidity bins (scaled by successive powers of $10^{2}$). The statistical and systematic uncertainties
are indicated by the error bars (too small to be seen on this scale)
and shaded bands, respectively. The points and horizontal error bars
indicate the \pt\ bin center and width, respectively.}
\label{fig:pp_xsec}
\end{figure}

The jet spectra, defined to be the average differential yield in a
given \pt\ bin, were constructed from a mixture of minimum-bias
(\PbPb\ only) and jet-triggered samples. In each \pt\ bin, the trigger
with the most events and that was more than 99\percent\ efficient for
that bin was used. The jet spectra were unfolded~\cite{Aad:2012vca} to
account for the \pt\ bin migration induced by the jet energy
resolution (JER) using a method based on the Singular Value
Decomposition~\cite{Hocker:1995kb}. The effects of the JER, which
receives contributions from both the detector response and UE
fluctuations, were evaluated by applying the same procedure to the MC
samples as was applied to the data and by matching the resulting
reconstructed jets and ``generator jets'' that are reconstructed from
final-state \textsc{Pythia} hadrons. For each pair, the \pt\ of the
generator and reconstructed jets were used to populate a detector
response matrix.  Separate response matrices were obtained for each
centrality interval.

The response matrix is generally diagonal, indicating that jets are
likely to be reconstructed in the same \pt\ bin as the generator
jets. The average \pt\ difference between reconstructed and generator
jets, is $\lesssim 1$\percent, independent of centrality. However, the
response distributions broaden at low \pt\ as the relative JER
increases due to the larger UE fluctuations. At $\pt=200$~\GeV, the
relative JER is approximately 10\percent\ and is independent of
centrality. However at $\pt=40$~\GeV, it varies from 20--40\percent\
between peripheral and central collisions. The unfolding is most
sensitive in this region and the range of jet \pt\ used in the
unfolding was chosen separately in each centrality interval to be as
low as possible while maintaining stability in the unfolding
procedure. The statistical covariance of each unfolded spectrum was
evaluated using the pseudo-experiment procedure described in
Ref.~\cite{Aad:2012vca}. Systematic uncertainties in the unfolding
procedure were evaluated by varying the choice of regularization
parameter used in the unfolding.
\begin{figure}[t]
  \centering
  \includegraphics[width=0.45\textwidth]{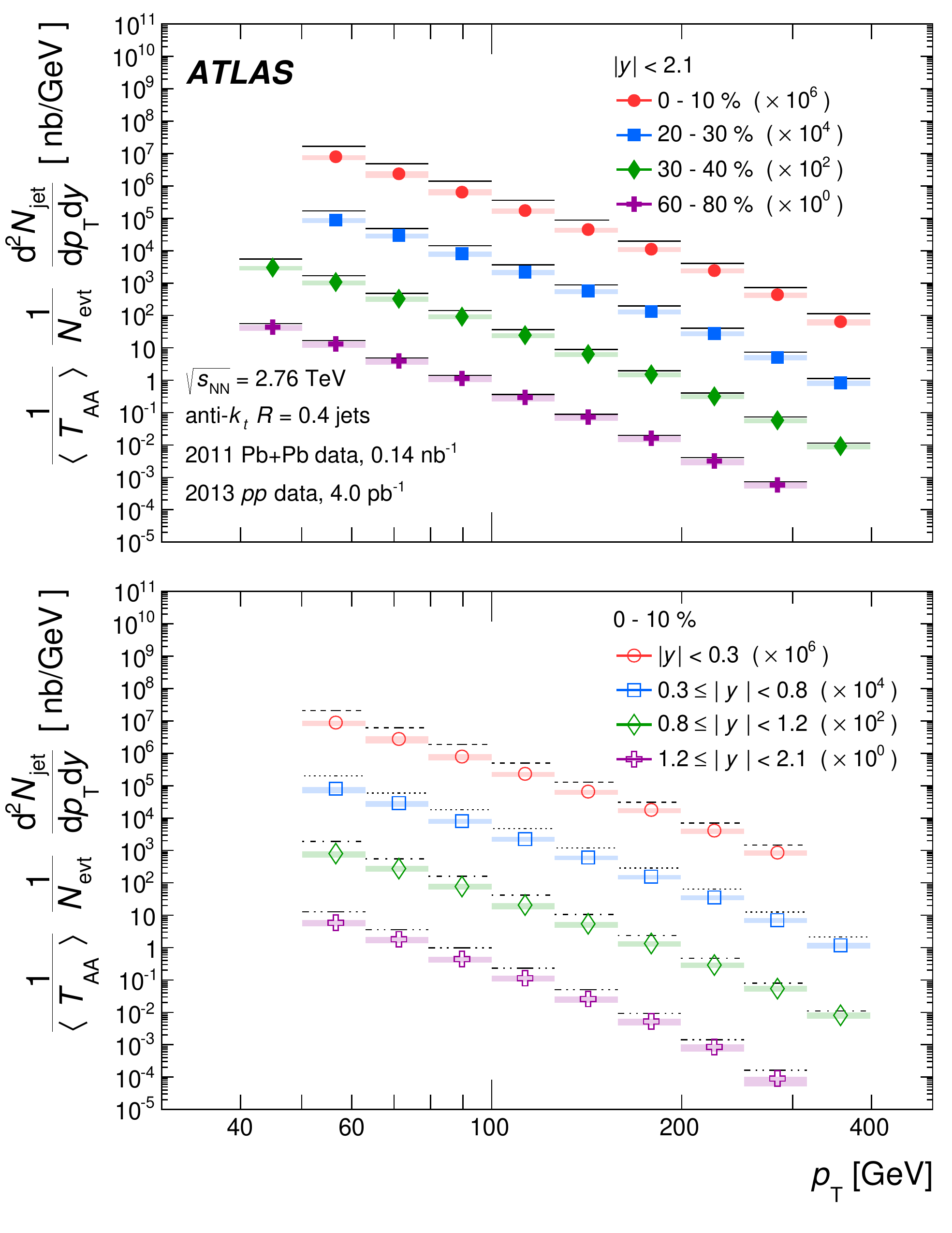}
  \caption{ The per-event jet yield in \PbPb\ collisions, multiplied
by $1/ \ATAA$, as a function of \pt\ (scaled by successive powers of
$10^{2}$). The upper panel shows the 0--2.1 rapidity range in
different centrality intervals. The lower panel shows the
0--10\percent\ centrality interval in different rapidity ranges. The
statistical and systematic uncertainties are indicated by the error
bars (too small to be seen on this scale) and shaded bands,
respectively. The points and horizontal error bars indicate the \pt\
bin center and width, respectively. The solid and dashed lines
represent the \pp\ jet cross section for the same rapidity interval
scaled by the same factor.}
 \label{fig:PbPb_xsec}
\end{figure}

The effects of any inefficiency in the jet reconstruction, including
inefficiency introduced by the UE jet rejection requirement, were
corrected for by a multiplicative correction applied after
unfolding. This factor, obtained from the MC sample, is unity for $\pT
> 100$~\GeV\ and reaches a maximum of 1.3 in the most central
collisions at the lowest \pT. For values larger than unity, an
uncertainty of 0.5\percent\ was assigned to this correction based on
the comparison of the jet reconstruction efficiency with respect to
track jets between the data and MC sample.

Uncertainties on the JER and JES have been evaluated using data-driven
techniques in \pp\ collisions~\cite{Aad:2011he,Aad:2012ag}. A
systematic uncertainty of 1.5\percent\ on the JES was assigned to
account for potential differences, not described by the MC
simulations, between the two data-taking periods. This value was
obtained by comparing the calorimetric response with respect to the
\pt\ of matched track jets in \pp\ and peripheral \PbPb\ collisions.
\begin{figure}[t] 
\centering
  \includegraphics[width=0.45\textwidth]{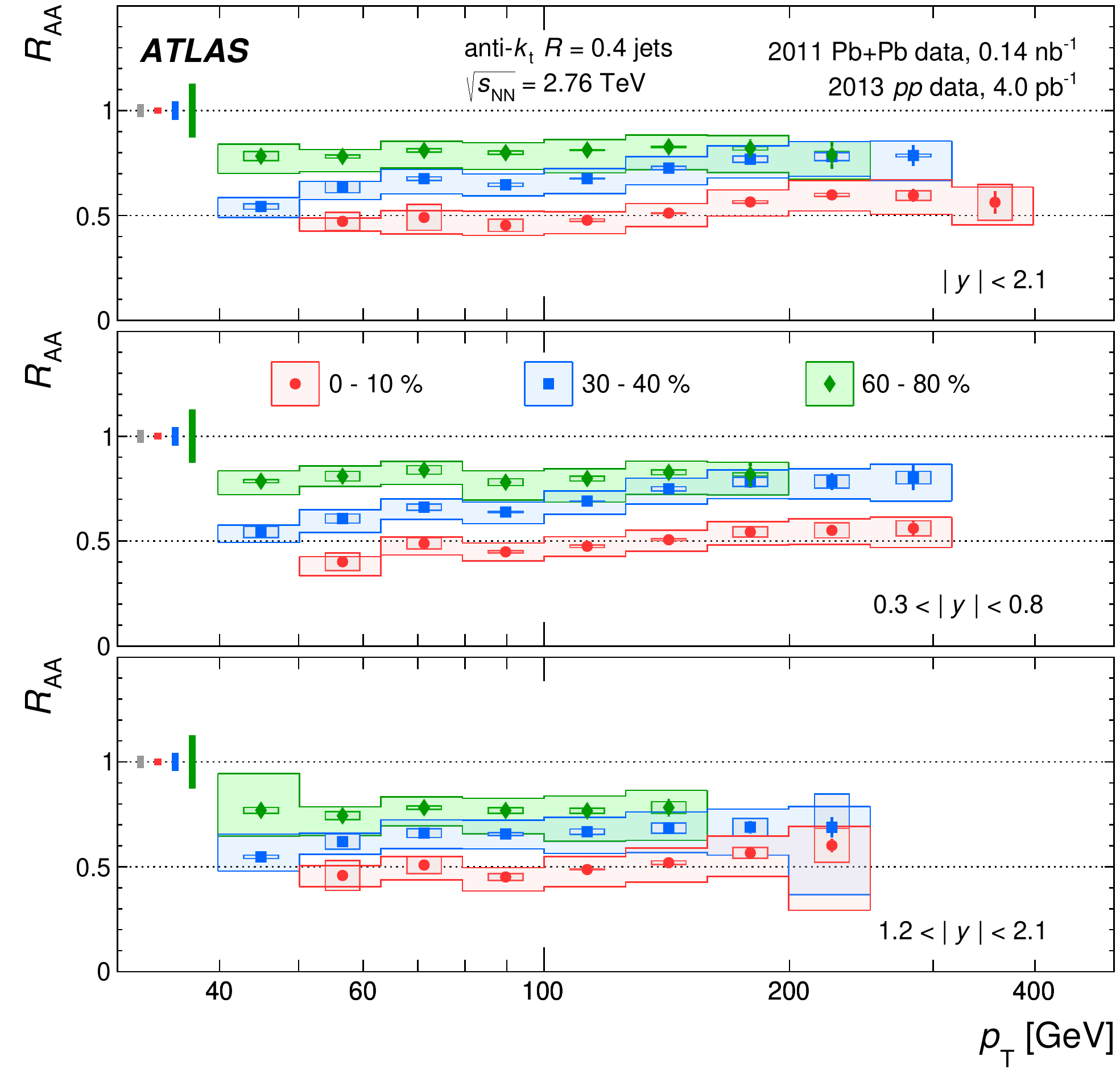}
  \caption{Jet \RAA\ as a function of \pt\ in different centrality
bins with each panel showing a different range in $|y|$. The
fractional luminosity and \ATAA\ uncertainties are indicated separately
as shaded boxes centered at one. The boxes, bands and error bars
indicate uncorrelated systematic, correlated systematic and
statistical uncertainties, respectively.}
  \label{fig:raa_pt_dep}
\end{figure}

A centrality-dependent uncertainty on the JES due to differences
between \pp\ and \PbPb\ in the partonic composition of jets and in
their fragmentation was estimated with the \textsc{Pyquen} sample.
The jet response in that sample was found to differ by up to
1\percent\ from that in the \textsc{Pythia} sample.  The magnitude of
this variation was checked with a similar study using track jets to
compare central and peripheral \PbPb\ data. The uncertainty was taken
to be 1\percent\ in the most central collisions with the uncertainty
decreasing in more peripheral collisions.

The impacts of the JER and JES uncertainties on the spectra were
assessed by constructing new response matrices with a systematically
varied relationship between the reconstructed and generator jet
kinematics and repeating the unfolding. Correlations in the JES and
JER uncertainties across the \pp\ and \PbPb\ samples were accounted
for in the propagation of the uncertainties to the \RAA.

Uncertainties on the \TAA\ and integrated luminosity affect the
overall normalization of the yields and thus are independent of jet
\pt\ and rapidity. The uncertainties on \ATAA\ vary between 1\percent\
and 10\percent\ in the most central and peripheral collisions,
respectively, with the full set of values given in
Table~\ref{tab:centrality_table}. The uncertainty on the integrated
luminosity is estimated to be \LumiUncert. It is determined, following
the same methodology as that detailed in Ref.~\cite{Aad:2013ucp}, from
a calibration of the luminosity scale derived from beam-separation
scans performed during the 2.76~\TeV\ operation of the LHC in 2013.

The total systematic uncertainty on the \pp\ cross sections is
dominated by the JES uncertainty, which is as large as 15\percent. For
the \PbPb\ jet yields this uncertainty is also dominant and in the
most central collisions is 22\percent. In the \RAA, much of this
uncertainty cancels. However the dominant contribution is due to the
JES in most centrality and rapidity intervals and is typically
10\percent. The uncertainties due to the unfolding are generally a few
per cent, but for some \pt\ values near the upper and lower limits
included in the measurement the contributions from this source are as
large as 15\percent. The contributions of the JER to the total
uncertainty on \RAA\ are less than 3\percent\ except in the most
central collisions at low \pt\ where they are as large as
10\percent. In the most peripheral bins the \ATAA\ uncertainties that
affect the overall normalization are the dominant contribution.

\begin{figure}[t]
\centering
\includegraphics[width=0.45\textwidth]{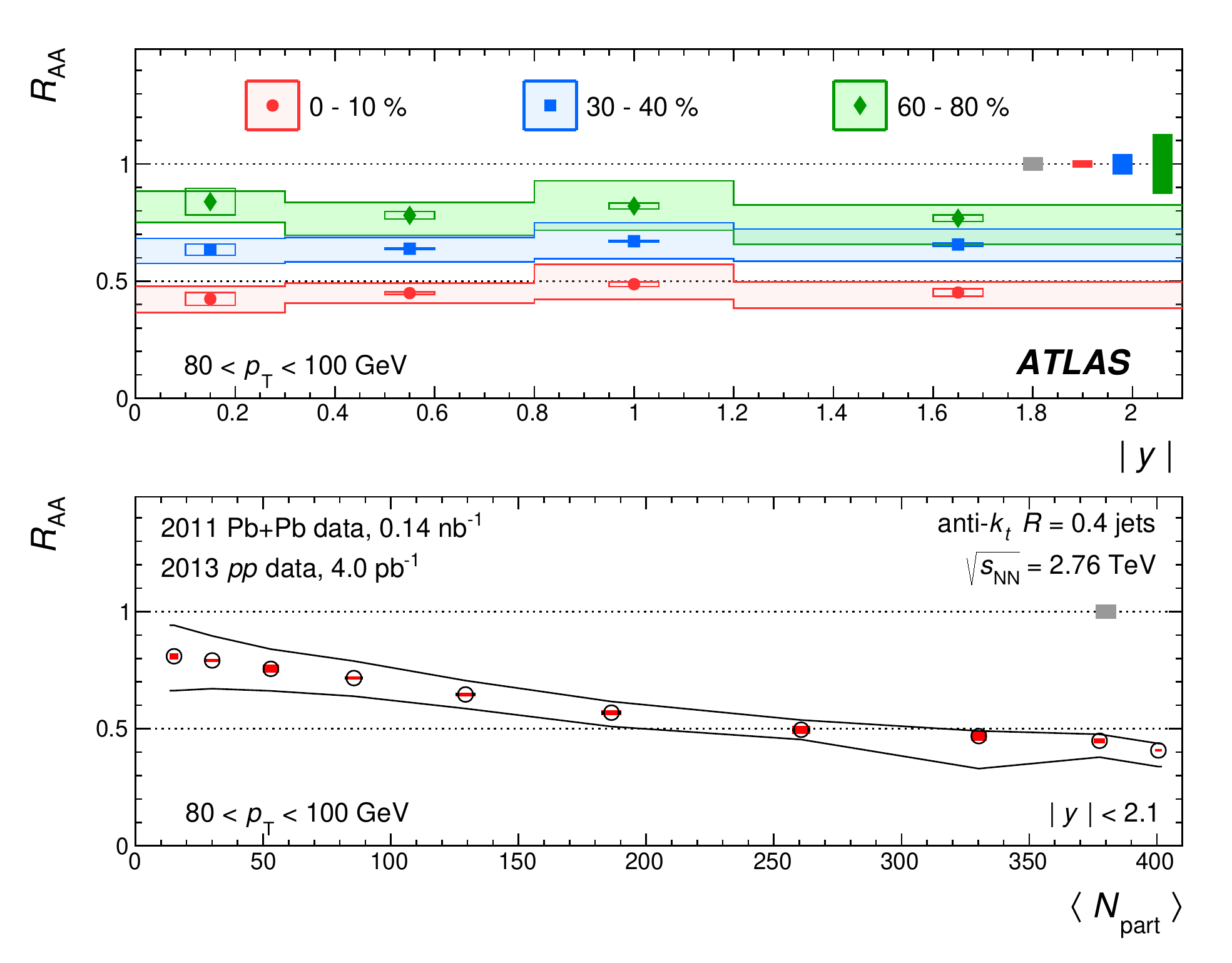}
\caption{The \RAA\ for jets with $ 80 < \pt < 100$~\GeV\ as a
function of \absy\ for different centrality bins (top) and as a
function of \ANpart\ for the $|y| < 2.1$ range (bottom). The fractional
luminosity and \ATAA\ uncertainties are indicated separately as shaded
boxes centered at one. The boxes, bands and error bars indicate
uncorrelated systematic, correlated systematic and statistical
uncertainties, respectively.}
  \label{fig:raa_y_npart_dep}
\end{figure}
The \pp\ differential jet cross sections are shown in
Fig.~\ref{fig:pp_xsec} for the following absolute rapidity ranges:
0--0.3, 0.3--0.8, 0.8--1.2, 1.2--2.1 and 0--2.1. These results are
consistent with a previous measurement with fewer
events~\cite{Aad:2013lpa}.  The differential per-event jet yield in
\PbPb\ collisions, multiplied by $1/\ATAA$, is shown in
Fig.~\ref{fig:PbPb_xsec}, in selected rapidity and centrality bins in
the lower and upper panels, respectively. The dashed lines represent
the \pp\ jet cross sections for that same rapidity bin; the jet
suppression is evidenced by the fact that the jet yields fall below
these lines.

The jet \RAA\ as a function of \pt\ is shown in
Fig.~\ref{fig:raa_pt_dep} for different ranges in collision centrality
and jet rapidity.  The \RAA\ is observed to increase weakly with \pt,
except in the most peripheral collisions. In the 0--10\percent\ and
$|y| < 2.1$ centrality and rapidity intervals, which have the smallest
statistical uncertainty, the \RAA\ is 0.47 at $\pt\sim55$ ~\GeV\ and
rises to 0.56 at $\pt\sim350$~\GeV.  These distributions were fit,
accounting for the point-wise correlations in the uncertainties, to
the functional form $a \ln (\pt) +b$. The slope parameter was found to
be significantly above zero in all but the most peripheral
collisions. The magnitude and weak increase of the \RAA\ in central
collisions are described quantitatively by recent theoretical
calculations ~\cite{He:2011pd,Kang:2014xsa}.  The results of this
measurement are consistent with measurements of the jet
central-to-peripheral ratio~\cite{Aad:2012vca}, although in those
measurements the uncertainties are too large to infer any significant
\pt\ dependence.

The rapidity dependence of the \RAA\ is shown in the top panel of
Fig.~\ref{fig:raa_y_npart_dep} for jets with $80 < \pt < 100$~\GeV\
for three centrality bins. The \RAA\ shows no significant rapidity
dependence over the \pt\ and rapidity ranges presented in this
measurement. The \ANpart\ dependence is shown in the bottom panel of
Fig.~\ref{fig:raa_y_npart_dep} for jets in the same \pt\ interval and
with $|y| < 2.1$. The \RAA\ decreases smoothly from the most
peripheral collisions (smallest \ANpart\ values) to central
collisions, where it reaches a minimal value of approximately 0.4 in
the most central 1\percent\ of collisions. A similar \ANpart\
dependence is observed for jets in different ranges of \pt\ and
rapidity.

In summary, this \papertype\ presents measurements of inclusive jet
production in \pp\ and \PbPb\ collisions over a wide range in \pt,
rapidity and centrality. The jet nuclear modification factor, \RAA,
obtained from these measurements shows a weak rise with \pt, with a
slope that varies with collision centrality.  No significant slope is
observed in the most peripheral collisions. The \RAA\ decreases
gradually with increasing \ANpart. At forward rapidity, the increasing
steepness of the jet production spectrum is expected to result in more
suppression of the jet yields. In this kinematic region, the
production is increasingly dominated by quark jets, which may lose
less energy than gluon jets~\cite{Baier:2000mf}. The observed lack of
rapidity dependence in the \RAA\ places constraints on relative energy
loss for quark and gluon jets in theoretical descriptions of jet
quenching.

We thank CERN for the very successful operation of the LHC, as well as
the support staff from our institutions without whom ATLAS could not
be operated efficiently.

We acknowledge the support of ANPCyT, Argentina; YerPhI, Armenia; ARC,
Australia; BMWFW and FWF, Austria; ANAS, Azerbaijan; SSTC, Belarus;
CNPq and FAPESP, Brazil; NSERC, NRC and CFI, Canada; CERN; CONICYT,
Chile; CAS, MOST and NSFC, China; COLCIENCIAS, Colombia; MSMT CR, MPO
CR and VSC CR, Czech Republic; DNRF, DNSRC and Lundbeck Foundation,
Denmark; EPLANET, ERC and NSRF, European Union; IN2P3-CNRS,
CEA-DSM/IRFU, France; GNSF, Georgia; BMBF, DFG, HGF, MPG and AvH
Foundation, Germany; GSRT and NSRF, Greece; ISF, MINERVA, GIF, I-CORE
and Benoziyo Center, Israel; INFN, Italy; MEXT and JSPS, Japan; CNRST,
Morocco; FOM and NWO, Netherlands; BRF and RCN, Norway; MNiSW and NCN,
Poland; GRICES and FCT, Portugal; MNE/IFA, Romania; MES of Russia and
ROSATOM, Russian Federation; JINR; MSTD, Serbia; MSSR, Slovakia; ARRS
and MIZ\v{S}, Slovenia; DST/NRF, South Africa; MINECO, Spain; SRC and
Wallenberg Foundation, Sweden; SER, SNSF and Cantons of Bern and
Geneva, Switzerland; NSC, Taiwan; TAEK, Turkey; STFC, the Royal
Society and Leverhulme Trust, United Kingdom; DOE and NSF, United
States of America.

The crucial computing support from all WLCG partners is acknowledged
gratefully, in particular from CERN and the ATLAS Tier-1 facilities at
TRIUMF (Canada), NDGF (Denmark, Norway, Sweden), CC-IN2P3 (France),
KIT/GridKA (Germany), INFN-CNAF (Italy), NL-T1 (Netherlands), PIC
(Spain), ASGC (Taiwan), RAL (UK) and BNL (USA) and in the Tier-2
facilities worldwide.

\bibliography{HION-2013-06}

\begin{thebibliography}{39}
\bibitem [{atl()}]{atlasfootnote}%
  \BibitemOpen
  \bibinfo {note} {{ ATLAS uses a right-handed coordinate system with its
  origin at the nominal interaction point (IP) in the center of the detector
  and the $z$-axis along the beam pipe. The $x$-axis points from the IP to the
  center of the LHC ring, and the $y$-axis points upward. Cylindrical
  coordinates $(r,\phi)$ are used in the transverse plane, $\phi$ being the
  azimuthal angle around the beam pipe. The pseudorapidity is defined in terms
  of the polar angle $\theta$ as $\eta=-\ln\tan(\theta/2)$.}}\BibitemShut
  {Stop}%
\end{thebibliography}
\bibliographystyle{apsrev4-1}
\clearpage
\onecolumngrid
\begin{flushleft}
{\Large The ATLAS Collaboration}

\bigskip

G.~Aad$^{\rm 84}$,
B.~Abbott$^{\rm 112}$,
J.~Abdallah$^{\rm 152}$,
S.~Abdel~Khalek$^{\rm 116}$,
O.~Abdinov$^{\rm 11}$,
R.~Aben$^{\rm 106}$,
B.~Abi$^{\rm 113}$,
M.~Abolins$^{\rm 89}$,
O.S.~AbouZeid$^{\rm 159}$,
H.~Abramowicz$^{\rm 154}$
% ...
\bigskip
\\
$^{1}$ Department of Physics, University of Adelaide, Adelaide, Australia\\
% ...
$^{11}$ Institute of Physics, Azerbaijan Academy of Sciences, Baku, Azerbaijan\\
% ...
$^{84}$ CPPM, Aix-Marseille Universit{\'e} and CNRS/IN2P3, Marseille, France\\
% ...
$^{89}$ Department of Physics and Astronomy, Michigan State University, East Lansing MI, United States of America\\
% ...
$^{106}$ Nikhef National Institute for Subatomic Physics and University of Amsterdam, Amsterdam, Netherlands\\
% ...
$^{112}$ Homer L. Dodge Department of Physics and Astronomy, University of Oklahoma, Norman OK, United States of America\\
$^{113}$ Department of Physics, Oklahoma State University, Stillwater OK, United States of America\\
% ...
$^{116}$ LAL, Universit{\'e} Paris-Sud and CNRS/IN2P3, Orsay, France\\
% ...
$^{152}$ Institute of Physics, Academia Sinica, Taipei, Taiwan\\
% ...
$^{154}$ Raymond and Beverly Sackler School of Physics and Astronomy, Tel Aviv University, Tel Aviv, Israel\\
% ...
$^{159}$ Department of Physics, University of Toronto, Toronto ON, Canada\\
% ...
\end{flushleft}

\end{document}